%% file: Motzkin_NQS.tex
\documentclass[aps,prx,reprint,preprintnumbers,amsmath,amssymb,superscriptaddress,nofootinbib,
floatfix]{revtex4-2}

\usepackage[colorlinks,bookmarks=true,citecolor=blue,linkcolor=red,urlcolor=blue]{hyperref}
\usepackage{graphicx}
\usepackage{braket}
\usepackage{ragged2e}
\usepackage{bm}
\usepackage{tikz}
\usepackage{ulem}
\usepackage{adjustbox}
\tikzset{every picture/.style={line width=0.75pt}}
\usepackage[dvipsnames]{xcolor}

\newcommand{\Motzkin}{\operatorname{Motzkin}}

\begin{document}
%\title{Exact Neural-Network Quantum State Representation of Motzkin States}
\title{Exact Neural-Network Representations of the Motzkin States}

\author{Runde Zha}\thanks{These authors contributed equally to this work.}
\affiliation{Beijing National Laboratory for Condensed Matter Physics, Institute of Physics,
Chinese Academy of Sciences, Beijing 100190, China}
\affiliation{School of Physical Sciences, University of Chinese Academy of Sciences, Beijing
100049, China}

\author{Yuntian Gu}\thanks{These authors contributed equally to this work.}
\affiliation{State Key Laboratory of General Artificial Intelligence, School of Intelligence
Science and Technology, Peking University}
\affiliation{ByteDance Seed, China}

\author{Chaohui Fan}\thanks{These authors contributed equally to this work.}
\affiliation{Beijing National Laboratory for Condensed Matter Physics, Institute of Physics,
Chinese Academy of Sciences, Beijing 100190, China}
\affiliation{School of Physical Sciences, University of Chinese Academy of Sciences, Beijing
100049, China}
\affiliation{ByteDance Seed, China}

\author{Jia-lin Chen}
\affiliation{Beijing National Laboratory for Condensed Matter Physics, Institute of Physics,
Chinese Academy of Sciences, Beijing 100190, China}
\affiliation{School of Physical Sciences, University of Chinese Academy of Sciences, Beijing
100049, China}

\author{Hai-Jun Liao}\email{navyphysics@iphy.ac.cn}
\affiliation{Beijing National Laboratory for Condensed Matter Physics, Institute of Physics,
Chinese Academy of Sciences, Beijing 100190, China}

\author{Tao Xiang}\email{txiang@iphy.ac.cn}
\affiliation{Beijing National Laboratory for Condensed Matter Physics, Institute of Physics,
Chinese Academy of Sciences, Beijing 100190, China}
\affiliation{School of Physical Sciences, University of Chinese Academy of Sciences, Beijing
100049, China}
\affiliation{Beijing Academy of Quantum Information Sciences, Beijing 100193, China}

\begin{abstract}
  Motzkin spin chains are paradigmatic frustration-free one-dimensional quantum systems whose ground states feature exactly solvable combinatorial structures and exotic, area-law-violating entanglement scaling. Specifically, colorless Motzkin states exhibit critical logarithmic entanglement divergence \(\log N\) with system size \(N\), while their colorful counterparts host supercritical sublinear \(\sqrt{N}\) entanglement growth.  Such unconventional entanglement behaviors place these states well beyond the expressive capability of standard matrix product states, which are fundamentally constrained by the entanglement area law. Here, we systematically construct exact, training-free neural-network representations for both colorless and colorful Motzkin states across four mainstream architectures, including recurrent, feedforward, convolutional, and transformer networks. Our core design leverages a causal prefix-sum module, implementable via recurrent updates, feedforward mappings, or masked attention layers, combined with position-selective rectified linear gates that enforce the Motzkin height constraints. For the colorful states, we further introduce a dedicated causal stack module that explicitly encodes the last-in-first-out color-matching rule.  Our results demonstrate that neural architectures can accurately capture highly non-trivial entanglement features inaccessible to conventional tensor networks, providing prototypic examples for benchmarking and a constructive design framework for future neural-network quantum state developments targeting strongly entangled quantum systems.

\end{abstract}

\maketitle

\section{Introduction}

Quantum entanglement is one of the most distinctive features of quantum mechanics, revealing correlations with no classical counterpart. In quantum many-body systems, it measures nonclassical correlations, constrains the complexity of quantum states, and distinguishes quantum phases beyond local order parameters. In one-dimensional gapped systems, ground-state entanglement is typically constrained by an area law~\cite{hastings2007area,eisert2010area}, while critical systems described by conformal field theory show only logarithmic violations~\cite{holzhey1994geometric, vidal2003entanglement,korepin2004universality,calabrese2004entanglement, calabrese2009entanglement}. 

Motzkin spin chains provide an exceptional exactly solvable setting beyond this standard picture. The original spin-1 Motzkin chain is frustration-free and critical, with a ground state given by an equal-weight superposition of Motzkin paths, namely constrained random walks or balanced-parenthesis configurations~\cite{bravyi2012criticality}. Its colored generalization gives the first rigorously solvable local spin-chain example with supercritical entanglement, where the half-chain entanglement entropy grows as \(\sqrt{N}\) with $N$ the chain length, parametrically faster than logarithmic critical scaling while still arising from local frustration-free interactions~\cite{movassagh2016supercritical}. Because this anomalous entanglement arises from simple local constraints and an explicitly known wave function, Motzkin states provide a controlled setting for benchmarking the quantum-state preparation and simulation protocols implemented on quantum computers and quantum simulators, as well as the representational power of tensor-network states, neural-network quantum states (NQS), and other variational wavefunctions.
 
This benchmark role has motivated recent work along several complementary directions. On the state-preparation side, digital dissipative protocols have been proposed for frustration-free gapless systems, including Motzkin-type spin chains~\cite{feldmeier2026dissipative}. On the simulation front, Rydberg-atom platforms offer a promising route to realizing Motzkin spin-chain constraints and dynamics in controllable atomic arrays~\cite{mukherjee2026rydberg}. From a representation perspective, Motzkin states have inspired exact tensor-network constructions, including holographic tensor-network descriptions of the Motzkin spin chain~\cite{alexander2021exact} and rainbow tensor networks for colorful Motzkin and Fredkin chains~\cite{alexander2019exact}. These developments reinforce the role of Motzkin states as analytically tractable but nontrivial testbeds for comparing how different physical platforms and classical ansatzes capture anomalous entanglement.

In recent years, NQS have emerged as highly expressive variational ansatzes for many-body wave functions~\cite{carleo2017solving}. Deep networks can efficiently encode broad classes of entangled states~\cite{gao2017efficient}, and close connections have been uncovered between neural-network states and tensor-network representations~\cite{chen2017equivalence,deng2021neural,han2018unsupervised, cheng2019tree,li2021boltzmann,fan2026disentangling}. The capabilities of NQS have been demonstrated in frustrated spin systems~\cite{choo2019two,nomura2021dirac,fan2026disentangling} and, more generally, in strongly correlated fermions. Early work adopted restricted Boltzmann machines to treat Hubbard models~\cite{nomura2017restricted}. Later architectures include autoregressive and recurrent networks for efficient sampling and sequential wave-function modeling~\cite{sharir2020deep,hibatAllah2020recurrent}, alongside determinant- and Pfaffian-based neural-network states that incorporate fermionic antisymmetry and pairing correlations~\cite{gauvinNdiaye2024mott,lange2025tj,chen2025pfaffian,choo2020fermionic,robledoMoreno2022fermionic,liu2024unifying,chen2025ml4ps}. Recently, autoregressive, transformer and convolutional backflow architectures have achieved high accuracy for doped Hubbard and related models, capturing stripe correlations and enabling large-scale benchmarks~\cite{ibarraGarciaPadilla2025autoregressive,gu2026hubbard,gu2026pareto}. At the same time, NQS have also been used as impurity solvers for quantum embedding~\cite{zhouyin2026impurity}. These developments highlight the strong representational capabilities of NQS and raise a natural question as to whether such frameworks can exactly encode critical and supercritical Motzkin states with entanglement properties that depart from standard area-law scaling.

 In this work, we resolve this question by constructing exact neural-network representations of both colorless and colorful Motzkin states. The construction is fully analytic and uses fixed weights, thereby eliminating the need for variational training. For the colorless state, a common height block computes prefix sums and excludes paths that violate nonnegativity or endpoint closure. For the colorful state, this block is supplemented by a stack, or equivalently by an attention pointer to the most recent unmatched up step, to enforce color matching. 

 The remainder of the paper is organized as follows. Section~\ref{sec:model} reviews the colorless and colorful Motzkin states, including their path constraints, parent Hamiltonians, and entanglement structure. Section~\ref{sec:colorless} presents the exact NQS construction for the colorless case, based on feed-forward, recurrent, and causal-attention implementations of the height rule. Section~\ref{sec:colorful} extends this construction to colorful Motzkin states by incorporating stack and attention-pointer blocks that perform color matching. Section~\ref{sec:discussion} analyzes how the parameter count scales with system size N for the various neural-network models and compares the resulting representations with tensor-network descriptions. Finally, Section~\ref{sec:summary} summarizes our findings and discusses implications and future directions.

\section{Motzkin States}
\label{sec:model}

\begin{figure}
\refstepcounter{figure} \label{fig:motzkin_path_examples}
\resizebox{0.78\columnwidth}{!}{\input{tikz_figs/motzkin_three_paths.tex}}
\smallskip
\begin{minipage}{0.92\columnwidth}
\justifying
\small \textbf{FIG.~\thefigure.}
\textbf{Examples of Motzkin-path constraints.} (a) A path that violates the height constraint by
crossing below the axis. (b) A legal colorful Motzkin path, where each down step matches the most
recent unmatched up step of the same color. (c) A path with the same height profile as in (b), but
with an illegal color mismatch.
\end{minipage}
\end{figure}

 Motzkin spin chains were introduced by Shor and coworkers as frustration-free one-dimensional models whose ground states are controlled by Motzkin paths~\cite{bravyi2012criticality}. Their colored generalization, developed by Movassagh and Shor, provided the first rigorously solvable local spin-chain example with supercritical entanglement~\cite{movassagh2016supercritical}. These models connect combinatorial path languages with quantum many-body physics and provide a useful testing ground for Hamiltonian complexity, entanglement structure, and variational representations.

Motzkin states are the ground states of Motzkin spin chains, with support defined by Motzkin
paths~\cite{bravyi2012criticality}. For a length-\(N\) spin-1 chain, the local basis
\(\{\ket{\uparrow},\ket{0},\ket{\downarrow}\}\) is identified with an up, flat, or down step with an increment
\begin{equation}
 x_i=
 \begin{cases}
 +1,& \mathrm{up \, step}=u,\\
 0,&  \mathrm{flat \, step}=0,\\
 -1,& \mathrm{down \, step}=d.
 \end{cases}
\end{equation}
A basis configuration \(X=(x_1,\ldots,x_N)\) defines prefix heights
\begin{equation}
 S_t=\sum_{i=1}^t x_i,\qquad S_0=0.
\end{equation}
The Motzkin constraint is
\begin{equation}
S_t\ge0\quad (t=1,\ldots,N-1),\qquad S_N=0,
\label{eq:motzkin_rule}
\end{equation}
namely the path never crosses below the axis and returns to zero at the endpoint. 

The colorless
Motzkin state is the equal-weight superposition of all configurations satisfying
Eq.~\eqref{eq:motzkin_rule},
\begin{equation}
\ket{\mathcal{M}_N}=
\frac{1}{\sqrt{|C_{0,0}(N)|}}
\sum_{s\in C_{0,0}(N)}\ket{s},
\label{eq:motzkin_state}
\end{equation}
where \(C_{0,0}(N)\) denotes the set of valid length-\(N\) Motzkin paths.

The colorful Motzkin state is obtained by assigning one of \(s\) colors, e.g., the $k$th color, to every up and down step,
\(\ket{u^k}\) and \(\ket{d^k}\), while flat steps remain colorless. In addition to the height rule
in Eq.~\eqref{eq:motzkin_rule}, every down step must match the color of the most recent unmatched
up step. Equivalently, the path obeys a last-in-first-out color-matching rule, as in balanced
colored parentheses; see Fig.~\ref{fig:motzkin_path_examples} for representative examples.

The normalized colorful Motzkin state is therefore
\begin{equation}
\ket{\mathcal{M}_{N,s}}=
\frac{1}{\sqrt{M_{N,s}}}
\sum_{w\in\Motzkin(N,s)}\ket{w},
\label{eq:colorful_state}
\end{equation}
where \(\Motzkin(N,s)\) is the set of color-matched Motzkin paths and
\(M_{N,s}=|\Motzkin(N,s)|\) is the number of all allowed paths.

\subsection{Parent Hamiltonians}

The Motzkin states arise as exact ground states of local frustration-free parent Hamiltonians. In the colorless case, the Hamiltonian is built from local equivalence moves~\cite{bravyi2012criticality}
\begin{equation}
00\leftrightarrow ud,
\qquad
0u\leftrightarrow u0,
\qquad
0d\leftrightarrow d0.
\label{eq:equivalence_moves}
\end{equation}
These moves make all paths in the same equivalence class have equal amplitude. The frustration-free Hamiltonian is
\begin{equation}
H =\Pi_{\rm b}+\sum_{j=1}^{N-1}\Pi_{j,j+1},
\end{equation}
where
\begin{eqnarray}
\Pi_{i,j}&=&\left(|\phi\rangle\langle \phi|
+|\psi_u\rangle\langle \psi_u|
+|\psi_d\rangle\langle \psi_d|\right)_{i,j},
\end{eqnarray}
with $\Pi_{\rm b}$ the boundary term
\begin{eqnarray}
\Pi_{\rm b}&=&|\downarrow\rangle\langle \downarrow|_1+|\uparrow\rangle\langle \uparrow|_N .
\end{eqnarray}
and
\begin{eqnarray}
\ket{\phi}&=&\frac{1}{\sqrt2}(\ket{\uparrow\downarrow}-\ket{00}),\\
\ket{\psi_u}&=&\frac{1}{\sqrt2}(\ket{\uparrow0}-\ket{0\uparrow}),\\
\ket{\psi_d}&=&\frac{1}{\sqrt2}(\ket{\downarrow0}-\ket{0\downarrow}).
\end{eqnarray}
The bulk projectors enforce equal amplitudes within each equivalence class, while the boundary terms remove all classes except \(C_{0,0}(N)\).  Eq.~\eqref{eq:motzkin_state} is therefore the unique zero-energy ground state.

For \(s\ge2\), each site carries \(\ket{0}\), \(\ket{u^k}\), or \(\ket{d^k}\), with \(k=1,\ldots,s\). The parent Hamiltonian~\cite{movassagh2016supercritical} adds local terms that enforce color matching:
\begin{equation}
H=\Pi_{\rm b}+\sum_{j=1}^{N-1}\Pi_{j,j+1}
+\sum_{j=1}^{N-1}\Pi_{j,j+1}^{\rm c},
\label{eq:colorful_parent_H}
\end{equation}
where
\begin{eqnarray}
  \Pi_{\rm b}&=&\sum_{k=1}^s\left( |d_k\rangle\langle d_k|_1+|u_k\rangle\langle u_k|_N\right),\\
  \Pi_{i,j}&=&\sum_{k=1}^s\left( |D_k\rangle\langle D_k|+ |U_k\rangle\langle U_k|+ |\phi_k\rangle\langle \phi_k|\right)_{i,j},\\
  \Pi_{i,j}^{\rm c}&=&\sum_{k\ne l}|u_k d_l\rangle\langle u_k d_l|_{i,j},
\end{eqnarray}
with
\begin{eqnarray}
\ket{D_k}&=&\frac{1}{\sqrt2}(\ket{d_k0}-\ket{0d_k}),\\
\ket{U_k}&=&\frac{1}{\sqrt2}(\ket{u_k0}-\ket{0u_k}),\\
\ket{\phi_k}&=&\frac{1}{\sqrt2}(\ket{u_kd_k}-\ket{00}).
\end{eqnarray}
The cross projectors remove locally mismatched adjacent pairs.  Together with the equivalence
moves in Eq.~\eqref{eq:equivalence_moves}, these local terms select
Eq.~\eqref{eq:colorful_state} as the unique zero-energy ground state. The case \(s=1\) reduces to
the colorless Hamiltonian because \(\Pi^{\rm c}\) vanishes.

\subsection{Entanglement entropy}

The unusual entanglement of Motzkin states follows from a Schmidt decomposition organized by the
height at the bipartition. For a half-chain cut, the left half can end at height \(h\), while the
right half must start from the same height and return to zero. In the colorless case, the Schmidt
sectors are labeled only by \(h\), and their weights are fixed by the number of partial Motzkin
paths with boundary height \(h\). This gives the logarithmic violation of the area law~\cite{bravyi2012criticality},
\begin{equation}
S_{s=1}\left(\frac{N}{2}\right)=\frac{1}{2}\ln N+\mathcal{O}(1).
\label{eq:colorless_entropy}
\end{equation}

The colored case is much more strongly entangled. If the height at the cut is \(h\), the \(h\)
unmatched up steps crossing the cut can carry \(s^h\) independent color strings. Since typical
heights scale as \(h\sim\sqrt{N}\), this color degeneracy produces a supercritical
\(\sqrt{N}\) contribution to the entropy~\cite{movassagh2016supercritical}. In natural logarithms,
the half-chain entropy behaves asymptotically as
\begin{equation}
S_{s\ge2}\left(\frac{N}{2}\right)
= 2\ln(s)\sqrt{\frac{\kappa N}{\pi}}
+\frac{1}{2}\ln(\pi\kappa N) + \mathcal{O}(1),
% +\gamma-\frac{1}{2},
\label{eq:colorful_entropy}
\end{equation}
where \(\kappa=\sqrt{s}/(2\sqrt{s}+1)\) . 

 Thus the colorless Motzkin chain is critical with logarithmic entanglement, whereas the colorful chain gives a rigorously solvable local spin chain with supercritical entanglement. It shows that locality and frustration-free alone do not guarantee weak entanglement.

 The neural network constructions below do not compute these entropies directly; instead, they reproduce the exact support and uniform amplitudes from which the Schmidt decompositions and entanglement scalings follow.

\section{Neural Representation of the Colorless Motzkin State}
\label{sec:colorless}

For the colorless Motzkin state, exact neural representation amounts to enforcing the path
constraints in Eq.~\eqref{eq:motzkin_rule} while assigning the same amplitude to every legal
configuration. This can be done in two steps: first compute each prefix height \(S_t\), and then
convert the nonnegativity and return-to-zero conditions into multiplicative gates. Based on this
principle, we construct four exact neural-network realizations for the colorless Motzkin state,
including recurrent neural network (RNN)~\cite{elman1990finding,hochreiter1997long,hibatAllah2020recurrent}, feedforward neural
network~(FNN)~\cite{rosenblatt1958perceptron,rumelhart1986learning,cybenko1989approximation}, convolutional neural network
(CNN)~\cite{lecun1998gradient,krizhevsky2012imagenet,he2016deep,liang2018convolutional}, and transformer~\cite{vaswani2017attention,devlin2019bert,zhang2023transformer}.

Throughout the constructions below, the activation function is defined by 
\begin{equation}
 \sigma(x)\equiv \mathrm{ReLU}(x)=\max(x,0) .
\end{equation}

 The network output must be divided by $\sqrt{M_{N,s}}$ to ensure proper normalization of the wavefunction. To illustrate the architecture, we present graphical representations for a chain of length $N=4$. However, the formalism applies generally to arbitrary $N$.  

\subsection{Recurrent Neural Network (RNN)}
\label{sec:RNN_colorless}

 The RNN architecture provides the most natural framework for realizing the Motzkin state in both the colorless and colored cases (the latter being discussed in Sec.~\ref{subsec:RNN_colorful}). Each neuron receives as input the local signal at the current site together with the hidden state from the preceding neuron, initialized by a parameter for the first neuron. This sequential dependence precludes parallel computation. However, it endows the RNN an intrinsic capacity for prefix summation and the associated constraint verification, both of which are essential for constructing the Motzkin state.

A recurrent network is illustrated schematically as
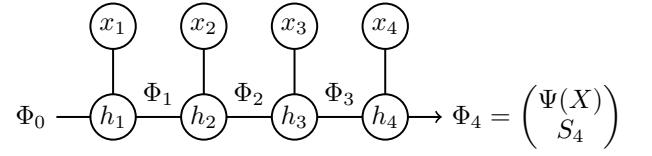
\[
\begin{tikzpicture}[every node/.style={scale=1},scale=0.6]
\draw (2,0) node {$x_1$} circle (0.5)
      (4,0) node {$x_2$} circle (0.5)
      (6,0) node {$x_3$} circle (0.5)
      (8,0) node {$x_4$} circle (0.5) ;

\draw (2,-2) node {$h_1$} circle (0.5)
      (4,-2) node {$h_2$} circle (0.5)
      (6,-2) node {$h_3$} circle (0.5)
      (8,-2) node {$h_4$} circle (0.5) ;

\draw (3,-1.5) node {$\Phi_1$}
      (5,-1.5) node {$\Phi_2$}
      (7,-1.5) node {$\Phi_3$} ;

\draw (2,-0.5) -- (2,-1.5);
\draw (4,-0.5) -- (4,-1.5);
\draw (6,-0.5) -- (6,-1.5);
\draw (8,-0.5) -- (8,-1.5);
\draw (0.75,-2) node [left] {$\Phi_0$} -- (1.5,-2);
\draw (2.5,-2) -- (3.5,-2);
\draw (4.5,-2) -- (5.5,-2);
\draw (6.5,-2) -- (7.5,-2);
\draw[->] (8.5,-2) -- (9.25,-2) [right] node  {$\Phi_4=\begin{pmatrix}
          \Psi(X)\\
          S_4
      \end{pmatrix}$} ;
\end{tikzpicture}
\]
with input configuration $X=(x_1,x_2,...,x_N)$.
It stores \(\Phi_i=(\psi_i,S_i)^T\), where $\Phi_i$ is the legality judgment of the input configuration until site $i$ and $S_i$ is the prefix sum. And the initialization parameter for neuron $h_1$ is \(\Phi_0=(1,0)^T\).  The operation in each neuron is
\begin{eqnarray}
\label{eq:RNN_colorless_S}
S_i & = &S_{i-1}+x_i,\\
\label{eq:RNN_colorless_psi}
\psi_i &= &\psi_{i-1}\cdot\sigma\left[1-h_i(S_i)\right],\\
\label{eq:RNN_colorless_activation}
h_i(S_i) &= &\sigma(-S_i)+\delta_{i,N}\cdot\sigma(S_i).
\end{eqnarray}
Thus \(\psi_i=1\) iff all constraints up to site \(i\) have been satisfied, and the network output
\(\Psi (X)=\psi_N\) is the exact support indicator.

\subsection{Feedforward Neural Network~(FNN)}
\label{subsec:FNN_colorless}

 The colorless Motzkin state can also be realized by generating the prefix sum through a linear transformation with a lower-triangular matrix $W$, and outputting the legality constraint $\Psi_i$ at each site via a designed activation function. The resulting wavefunction $\Psi(X)$ is a product of the local amplitudes $\psi_i$, which naturally yields a Motzkin state with equal-amplitude superposition over all legal configurations. 

The FNN representation of the colorless Motzkin state is schematically depicted as
 
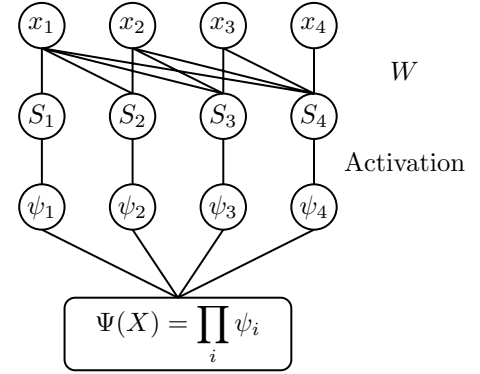
\[
\begin{tikzpicture}[every node/.style={scale=1},scale=0.6]
\draw (0,0) node {$x_1$} circle (0.5)
      (2,0) node {$x_2$} circle (0.5)
      (4,0) node {$x_3$} circle (0.5)
      (6,0) node {$x_4$} circle (0.5) ;

\draw (0,-2) node {$S_1$} circle (0.5)
      (2,-2) node {$S_2$} circle (0.5)
      (4,-2) node {$S_3$} circle (0.5)
      (6,-2) node {$S_4$} circle (0.5) ;

\draw (0,-4) node {$\psi_1$} circle (0.5)
      (2,-4) node {$\psi_2$} circle (0.5)
      (4,-4) node {$\psi_3$} circle (0.5)
      (6,-4) node {$\psi_4$} circle (0.5) ;

\draw (8,-1) node {$W$}
      (8,-3) node {Activation} ;

\draw[rounded corners] (5.5,-7.6) rectangle (0.5,-6) ;
\draw (3,-6.8) node {$\Psi (X) = \displaystyle \prod_i \psi_i  $} ;

\draw (0,-0.5) -- (0, -1.5) (0,-0.5) -- (2,-1.5) (0,-0.5) -- (4,-1.5) (0,-0.5) -- (6,-1.5)
      (2,-0.5) -- (2,-1.5) (2,-0.5) -- (4,-1.5) (2,-0.5) -- (6,-1.5)
      (4,-0.5) -- (4,-1.5) (4,-0.5) -- (6,-1.5)
      (6,-0.5) -- (6,-1.5) ;

\draw (0,-2.5) -- (0, -3.5) (2,-2.5) -- (2,-3.5) (4,-2.5) -- (4,-3.5) (6,-2.5) -- (6,-3.5) ;

\draw (0,-4.5) -- (3, -6) (2,-4.5) -- (3,-6) (4,-4.5) -- (3,-6) (6,-4.5) -- (3,-6) ;

\end{tikzpicture}
\]

 Starting from the input configuration $X=(x_1,x_2,\ldots, x_N)$, this network uses a lower-triangular matrix 
 \begin{equation}
W_{ij}=\begin{cases}
1,&j\le i,\\
0,&j>i ,
\end{cases}
\label{eq:prefix_sum}
\end{equation}
 to map the configuration to a sequence of prefix sums
 \begin{equation}
    S_i=\sum_j W_{ij}x_j=\sum_{j=1}^{i}{x_j} .
    \label{eq:linear_trans_FNN}
\end{equation}
While the linear transformation layer $W$ is fully connected, only solid lines corresponding to nonzero entries $W_{ij}$ are plotted for this particular construction.

Each \(S_i\) is subsequently fed into an activation function
\begin{equation}
   \psi_i =\sigma\left[ 1-\sigma(-S_i)-\delta_{i,N}\cdot\sigma(S_i) \right] .
\label{eq:FNN_gate}
\end{equation}
Since all \(S_i\) take integer values, the outputs satisfy \(\psi_i=1\) iff \(S_i\ge0\)
for all \(i<N\) otherwise 0, and \(\psi_N=1\) iff \(S_N=0\) otherwise 0. 

The final wavefunction is a product of all $\psi_i$
\begin{equation}
\Psi(X)=\prod_{i=1}^N\psi_i(X)
\label{eq:FNN_colorless_production}
\end{equation}
 which constitutes an exact neural network representation of the colorless Motzkin state. 

\subsection{Convolutional Neural Network (CNN)}

 In the FNN construction, a linear transformation $W$ is introduced to compute the prefix sums. The same objective can be achieved with a CNN. In the convolutional realization, $W$ is replaced by a convolutional layer that introduces $N-1$ ancilla nodes and generates the prefix sums sequentially. For example, $S_1$ is produced as
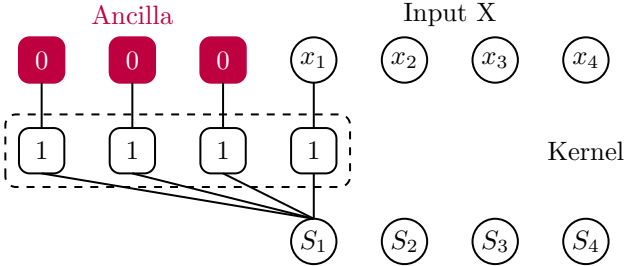
\[
\begin{tikzpicture}[every node/.style={scale=1},scale=0.6]
\draw (-2,1) node {\textcolor{purple}{Ancilla}}
      (5,1) node {Input X};
\draw[rounded corners,draw=purple,fill=purple] (-0.5,-0.5) rectangle (0.5,0.5) ;
\draw[rounded corners,draw=purple,fill=purple] (-2.5,-0.5) rectangle (-1.5,0.5) ;
\draw[rounded corners,draw=purple,fill=purple] (-4.5,-0.5) rectangle (-3.5,0.5) ;
\draw (0,0) node {\textcolor{white}{0}}
      (-2,0) node {\textcolor{white}{0}}
      (-4,0) node {\textcolor{white}{0}};
\draw (2,0) node {$x_1$} circle (0.5)
      (4,0) node {$x_2$} circle (0.5)
      (6,0) node {$x_3$} circle (0.5)
      (8,0) node {$x_4$} circle (0.5) ;
\draw (2,-4) node {$S_1$} circle (0.5)
      (4,-4) node {$S_2$} circle (0.5)
      (6,-4) node {$S_3$} circle (0.5)
      (8,-4) node {$S_4$} circle (0.5) ;
\draw[rounded corners, dashed] (-4.8,-2.8) rectangle (2.8,-1.2) ;
\draw[rounded corners] (-4.5,-2.5) rectangle (-3.5,-1.5) ;
\draw[rounded corners] (-2.5,-2.5)  rectangle (-1.5,-1.5) ;
\draw[rounded corners] (-0.5,-2.5)  rectangle (0.5,-1.5) ;
\draw[rounded corners] (1.5,-2.5)  rectangle (2.5,-1.5) ;
\draw (-2,-2) node {1}
      (0,-2) node {1}
      (2,-2) node {1}
      (-4,-2) node {1} ;
\draw (8,-2) node {Kernel};
\draw[] (-4,-0.5) -- (-4,-1.5);
\draw[] (-2,-0.5) -- (-2,-1.5);
\draw[] (0,-0.5) -- (0,-1.5);
\draw[] (2,-0.5) -- (2,-1.5);
\draw[] (0,-2.5) -- (2,-3.5);
\draw[] (2,-2.5) -- (2,-3.5);
\draw[] (-4,-2.5) -- (2,-3.5);
\draw[] (-2,-2.5) -- (2,-3.5);

\end{tikzpicture}
\]
Similarly, $S_3$ is obtained from 
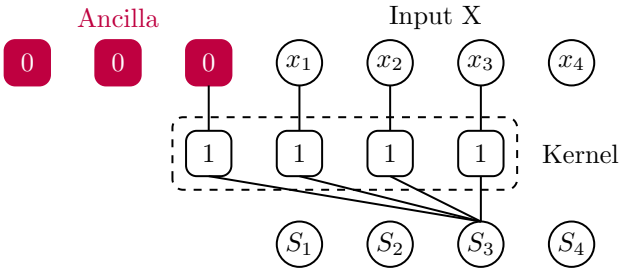
\[
\begin{tikzpicture}[every node/.style={scale=1},scale=0.6]
\draw (-2,1) node {\textcolor{purple}{Ancilla}}
      (5,1) node {Input X};
\draw[rounded corners,draw=purple,fill=purple] (-0.5,-0.5) rectangle (0.5,0.5) ;
\draw[rounded corners,draw=purple,fill=purple] (-2.5,-0.5) rectangle (-1.5,0.5) ;
\draw[rounded corners,draw=purple,fill=purple] (-4.5,-0.5) rectangle (-3.5,0.5) ;
\draw (0,0) node {\textcolor{white}{0}}
      (-2,0) node {\textcolor{white}{0}}
      (-4,0) node {\textcolor{white}{0}};
\draw (2,0) node {$x_1$} circle (0.5)
      (4,0) node {$x_2$} circle (0.5)
      (6,0) node {$x_3$} circle (0.5)
      (8,0) node {$x_4$} circle (0.5) ;
\draw (2,-4) node {$S_1$} circle (0.5)
      (4,-4) node {$S_2$} circle (0.5)
      (6,-4) node {$S_3$} circle (0.5)
      (8,-4) node {$S_4$} circle (0.5) ;
\draw[rounded corners, dashed] (-0.8,-2.8) rectangle (6.8,-1.2) ;
\draw[rounded corners] (-0.5,-2.5) rectangle (0.5,-1.5) ;
\draw[rounded corners] (1.5,-2.5)  rectangle (2.5,-1.5) ;
\draw[rounded corners] (3.5,-2.5)  rectangle (4.5,-1.5) ;
\draw[rounded corners] (5.5,-2.5)  rectangle (6.5,-1.5) ;
\draw (2,-2) node {1}
      (4,-2) node {1}
      (6,-2) node {1}
      (0,-2) node {1} ;
\draw (8.2,-2) node {Kernel};
\draw[] (0,-0.5) -- (0,-1.5);
\draw[] (2,-0.5) -- (2,-1.5);
\draw[] (4,-0.5) -- (4,-1.5);
\draw[] (6,-0.5) -- (6,-1.5);
\draw[] (0,-2.5) -- (6,-3.5);
\draw[] (2,-2.5) -- (6,-3.5);
\draw[] (4,-2.5) -- (6,-3.5);
\draw[] (6,-2.5) -- (6,-3.5);

\end{tikzpicture}
\]
and the remaining prefix sums follow in the same manner. Here, the $N-1$ ancilla are initialized to 0 and placed ahead of the input array $X$, so that the full input configuration $x_i$ comprises $2N-1$ nodes, with $x_i=0$ for $i\le 0$ and the physical input otherwise. A uniform convolution kernel 
\begin{equation}
    {\rm Kernel}_i=1,\qquad  (1\le i\le N) 
    \label{eq:kernel}
\end{equation}
 is then applied, where the kernel window acting on $x_i$ covers the $N$ nodes $\{x_j;\, j=i-N+1,\dots,i\}$ counting backwards from $x_i$. This convolution yields
\begin{equation}
    S_i =\sum_{j=1}^N{{\rm Kernel}_j \cdot x_{i+j-N}}= \sum_{j=1}^i x_{j}\,.
\end{equation}
 which reproduces precisely the triangular map of Eq.~\eqref{eq:prefix_sum}.  Finally, the gates of Eq.~\eqref{eq:FNN_gate}, together with the production rule of Eq.~\eqref{eq:FNN_colorless_production} complete the CNN representation of the Motzkin state.

\subsection{Transformer}
\label{sec:attention}

 The transformer, which is the most prevalent neural network architecture to date, owes its success to its all-to-all connectivity, which captures correlations at all scales. The Motzkin state can also be represented by a single-layer transformer, as depicted in
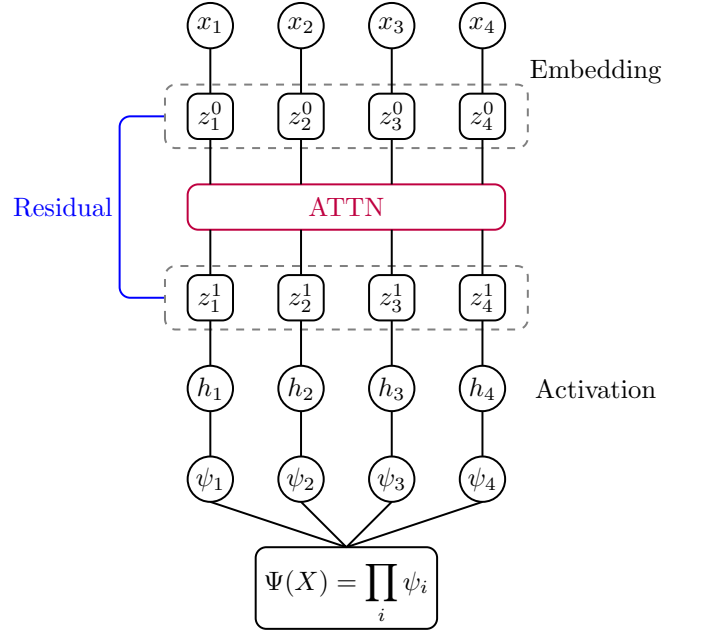
\[
\begin{tikzpicture}[every node/.style={scale=1},scale=0.6]
\draw (0,0) node {$x_1$} circle (0.5)
      (2,0) node {$x_2$} circle (0.5)
      (4,0) node {$x_3$} circle (0.5)
      (6,0) node {$x_4$} circle (0.5) ;
\draw (8.5,-1) node {Embedding}
      (8.5,-8) node {Activation};
\draw[draw=blue,rounded corners] (-1,-2)--(-2,-2)--(-2,-6)--(-1,-6);
\draw (-3.25,-4) node {\textcolor{blue}{Residual}};
\draw[dashed,rounded corners,draw=gray] (-1,-2.7) rectangle (7,-1.3);
\draw[dashed,rounded corners,draw=gray] (-1,-6.7) rectangle (7,-5.3);

\draw (0,-2) node {$z_1^{0}$}
      (2,-2) node {$z_2^0$}
      (4,-2) node {$z_3^0$}
      (6,-2) node {$z_4^0$} ;
\draw (0,-6) node {{$z_1^1$}}
      (2,-6) node {$z_2^1$}
      (4,-6) node {$z_3^1$}
      (6,-6) node {$z_4^1$} ;

\draw [rounded corners] (-0.5,-2.5) rectangle (0.5,-1.5);
\draw [rounded corners]  (1.5,-2.5) rectangle (2.5,-1.5);
\draw [rounded corners]  (3.5,-2.5) rectangle (4.5,-1.5);
\draw [rounded corners]  (5.5,-2.5) rectangle (6.5,-1.5);
\draw [rounded corners] (-0.5,-6.5) rectangle (0.5,-5.5);
\draw [rounded corners]  (1.5,-6.5) rectangle (2.5,-5.5);
\draw [rounded corners]  (3.5,-6.5) rectangle (4.5,-5.5);
\draw [rounded corners]  (5.5,-6.5) rectangle (6.5,-5.5);
\draw [rounded corners,draw=purple] (-0.5,-4.5)  rectangle (6.5,-3.5);
\draw (3,-4) node {\textcolor{purple}{ATTN}};
\draw (0, -0.5) -- (0,-1.5) (0, -2.5) -- (0,-3.5) (0,-4.5)--(0,-5.5)
      (2, -0.5) -- (2,-1.5) (2, -2.5) -- (2,-3.5) (2,-4.5)--(2,-5.5)
      (4, -0.5) -- (4,-1.5) (4, -2.5) -- (4,-3.5) (4,-4.5)--(4,-5.5)
      (6, -0.5) -- (6,-1.5) (6, -2.5) -- (6,-3.5) (6,-4.5)--(6,-5.5);
\draw (0,-8) node {{$h_1$}} circle (0.5);
\draw (2,-8) node {$h_2$} circle (0.5)
      (4,-8) node {$h_3$} circle (0.5)
      (6,-8) node {$h_4$} circle (0.5) ;
\draw (0,-10) node {{$\psi_1$}} circle (0.5);
\draw (2,-10) node {$\psi_2$} circle (0.5)
      (4,-10) node {$\psi_3$} circle (0.5)
      (6,-10) node {$\psi_4$} circle (0.5) ;
\draw (2,-6.5)--(2,-7.5) 
      (4,-6.5)--(4,-7.5) 
      (6,-6.5)--(6,-7.5);
\draw (0,-6.5)--(0,-7.5) ;
\draw (2,-8.5)--(2,-9.5) 
      (4,-8.5)--(4,-9.5) 
      (6,-8.5)--(6,-9.5);
\draw (0,-8.5)--(0,-9.5) ;
\draw[rounded corners] (1,-13.3) rectangle (5,-11.5);
\draw (3,-12.5) node{$\Psi(X)=\displaystyle \prod_i {\psi_i}$};
\draw (0,-10.5)--(3,-11.5) (2,-10.5)--(3,-11.5) (4,-10.5)--(3,-11.5) (6,-10.5)--(3,-11.5);
\end{tikzpicture}
\]

Although it seems that the transformer construction is more complicated in the figure, the attention block works similarly to $W$ matrix in FNN and the convolution kernel and outputs a prefix average for every site. The advantage of the transformer is that it can be most easily generalized to other models we may further investigate, e.g., some perturbed situation of the Motzkin model.

 This transformer architecture consists of an attention block mapping $z^0$ to $z^1$, followed by an activation block mapping $z^1$ to $\psi$. Below, we describe each block in detail.

\subsubsection{Attention block}

In this transformer construction, each site is first embedded as 
\begin{equation}
z_i^0=(x_i,e_i,0)^T\in\mathbb{R}^{N+2},
\end{equation}
where \(e_i\) is the \(i\)th standard basis vector in position space, i.e. 
\begin{equation}
    (e_i)_j=\delta_{i,j},
\end{equation}
and the last component is left empty to receive the attention output.   

 The attention block uses the projection matrices $W_Q,W_K,W_V$ to map the input $z^0$ into queries, keys, and values, respectively. The queries and keys are combined into Logit scores, which a softmax converts into attention weights $\alpha(t,i)$. These weights then multiply the values, and the resulting output, added to the residual (here $z^0$ itself), yields the $z^1$ layer. Graphically, this block is
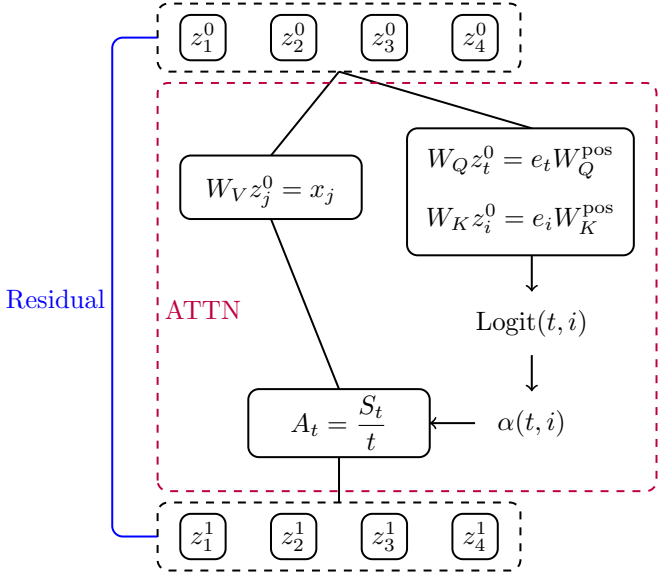
\[
\begin{tikzpicture}[every node/.style={scale=1},scale=0.6]
\draw [rounded corners] (-0.5,-0.5) rectangle (0.5,0.5);
\draw [rounded corners]  (1.5,-0.5) rectangle (2.5,0.5);
\draw [rounded corners]  (3.5,-0.5) rectangle (4.5,0.5);
\draw [rounded corners]  (5.5,-0.5) rectangle (6.5,0.5);
\draw (0,0) node {$z_1^0$}
      (2,0) node {$z_2^0$}
      (4,0) node {$z_3^0$}
      (6,0) node {$z_4^0$} ;
\draw[dashed, rounded corners] (-1,0.75) rectangle (7,-0.75) ;

\draw[] (3,-0.75)--(1.5,-2.6) (3,-0.75)--(7.25,-2) ;
\draw[dashed,draw=purple,rounded corners] (-1,-10) rectangle (10,-1);
\draw (0,-6) node {\textcolor{purple}{ATTN}};
\draw[rounded corners] (4.5,-4.8) rectangle (9.5,-2);
\draw (7,-2.75) node {$W_Qz_t^0=e_tW_Q^{\mathrm{pos}}$}
      (7,-4) node{$W_Kz_i^0=e_iW_K^{\mathrm{pos}}$} ;
\draw[rounded corners] (-0.5,-4) rectangle (3.5,-2.6);
\draw (1.5,-3.4) node {$W_Vz_j^0=x_j$};

\draw[->] (7.25,-4.8)--(7.25,-5.6) ;
\draw (7.25,-6.3) node {$\mathrm{Logit}(t,i)$};
\draw[->] (7.25,-7)--(7.25,-7.8) ;
\draw[->] (7.25,-8.5) node {$\alpha(t,i)$} (6, -8.5) --(5,-8.5);

\draw[rounded corners] (3,-8.5) node {$A_t =\displaystyle \frac{S_t}{t}$}  (1,-7.75) rectangle (5,-9.25) ;

\draw (1.5,-4) --(3,-7.75) ;
\draw[] (3,-9.25)--(3,-10.25);
\draw[dashed, rounded corners] (-1,-10.25) rectangle (7,-11.75) ;

\draw [rounded corners] (-0.5,-11.5) rectangle (0.5,-10.5);
\draw [rounded corners]  (1.5,-11.5) rectangle (2.5,-10.5);
\draw [rounded corners]  (3.5,-11.5) rectangle (4.5,-10.5);
\draw [rounded corners]  (5.5,-11.5) rectangle (6.5,-10.5);
\draw (0,-11) node {$z_1^1$}
      (2,-11) node {$z_2^1$}
      (4,-11) node {$z_3^1$}
      (6,-11) node {$z_4^1$} ;

\draw[rounded corners, draw=blue] (-1,0)--(-2,0)--(-2,-11)--(-1,-11);
\draw (-3.25,-5.75) node {\textcolor{blue}{Residual}};
\end{tikzpicture}
\]

To implement a causally masked attention layer, we take the positional key matrix to be the identity, \(W_K^{\rm pos}=I_N\), and the positional query matrix to be 
\begin{equation}
W_Q^{\rm pos}(t,i)=
\begin{cases}
0,&i\le t,\\
-\infty,&i>t,
\end{cases}
\label{eq:attn_mask}
\end{equation}
or, equivalently, a finite penalty \(-C\) in the limit  \(C\to\infty\). Setting the first and the last rows and columns of $W_{Q/K}$ to zero so that $W_{Q/K}$ acts only on the position subspace and decouples the physical variable and the empty slot from the attention scores, giving 
\begin{equation}
    W_Qz_t^0 \cdot W_Kz_i^0 = e_tW_Q^{\mathrm{pos}} \cdot e_iW_K^{\mathrm{pos}} 
\end{equation}
The logit score between site \(t\) and site \(i\) is then
\begin{eqnarray}
    \mathrm{Logit}(t,i)= W_Qz_t^0 \cdot W_Kz_i^0
    %\notag\\&=&
 = \begin{cases}
0,&i\le t,\\
-\infty,&i>t.
\end{cases}
\end{eqnarray}
The softmax therefore yields exact causal weights
\begin{equation}
\alpha_{t,i}=\frac{\exp(\mathrm{Logit}(t,i))}{\sum_j{\exp (\mathrm{Logit}(t,j)}) }=\begin{cases}
1/t,&i\le t,\\
0,&i>t,
\end{cases}
\label{eq:attn_alpha}
\end{equation}
 and choosing the value matrix $W_V$ to extract the physical variable, i.e.
 \begin{eqnarray}
 (W_V)_{ij}&=&\begin{cases}
     1 & i=N+2,\, j=1 ,\\
     0 & \mathrm{otherwise} ,
 \end{cases}    \\
 W_Vz_i^0&=&(0,0,\dots,0,x_i)^T ,
 \end{eqnarray}
the attention output at site $t$ becomes the prefix average
\begin{equation}
A_t=\sum_{i=1}^N\alpha_{t,i}x_i=\frac{1}{t}\sum_{i=1}^t x_i=\frac{S_t}{t}.
\label{eq:prefix_avg}
\end{equation}

After the residual connection, the intermediate vector reads
\begin{equation}
z_t^1= (x_t,e_t,S_t/t)^T.
\end{equation}
i.e., the first $N+1$ components in $z_t^1$ are inherited unchanged from $z_t^0$, while the last component  $S_t/t$ is supplied by the attention block and equals the prefix average at site $t$. In numerical implementations, one may replace \(-\infty\) by a large negative constant \(-C\). The exact limit is recovered by the standard hard causal mask or by the \(C\to\infty\) limit.

\subsubsection{Activation}
\label{sec:relu}

Starting from \(z_t^1=(x_t,e_t,\bar{S}_t)^T\), obtained from the attention block with a residual connection, we define the hidden unit at site \(t\) as
\begin{eqnarray}
h_t(z_t^1)&=&\sigma\left(-t\bar{S}_t\right)+\sigma(1-N+t)\sigma\left(t\bar{S}_t\right) \nonumber \\
&=&\sigma\left(-S_t\right)+\delta_{t,N}\sigma\left(S_t\right).
\label{eq:height_hidden}
\end{eqnarray}
The corresponding site gate is
\begin{equation}
\psi_t=\sigma\left[1- h_t(z_t^1)\right],
\label{eq:height_gate}
\end{equation}
which can be written explicitly as
\begin{equation}
\psi_t=
\begin{cases}
\sigma\left[1-\sigma(-S_t)\right],&t<N,\\
\sigma\left[1-\sigma(-S_N)-\sigma(S_N)\right],&t=N.
\end{cases}
\label{eq:abs-relu}
\end{equation}
 Since \(S_t\) is integer-valued, \(\psi_t=1\) iff the corresponding Motzkin height constraint is satisfied, whereas \(\psi_t=0\) otherwise. Thus, the product of all site gates enforces the Motzkin constraints and yields the Motzkin-state amplitude.

\section{Neural Representation of the Colorful Motzkin States}
\label{sec:colorful}

  Colorless neural network constructions demonstrate that height constraints can be exactly enforced via neural gates. In contrast, the legality of colorful Motzkin states depends not only on path height, but also on the sequential order of color opening and closing operations. Characterizing valid configurations thus requires tracking the stack of unmatched colors and enforcing the corresponding last-in-first-out matching rule. In this section, we construct neural networks that implement this stack-based memory mechanism while maintaining uniform amplitudes across all valid colorful Motzkin paths.

For \(s\ge2\), each increment \(x_i\) is specified by a height increment \(\Delta_i\) and a color label \(c_i\):
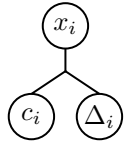
\[
\begin{tikzpicture}[every node/.style={scale=1},scale=0.6]
\draw (0.0,0) node {$x_i$} circle(0.5) ;
\draw (0.0,-0.5) -- (0.0,-1) -- (-0.75,-1.5) (0.0,-1) -- (0.75,-1.5) ;
\draw (-0.75,-2) node {$c_i$} circle(0.5) ;
\draw (0.75,-2) node {$\Delta_i$} circle(0.5) ;
\end{tikzpicture}
\]
The height increment $$\Delta_i$$ is defined independently of the color label via
\begin{equation}
\Delta_i=
\begin{cases}
+1,& x_i=u^k,\\
0,& x_i=0,\\
-1,& x_i=d^k,
\end{cases}
\end{equation}
while the color label is set to \(c_i = k\) for both \(u^k\) and \(d^k\). 

Valid colored equivalence moves are given by the pairwise local transformations~\cite{movassagh2016supercritical,menon2024motzkin}
\begin{equation}
0d^k\leftrightarrow d^k0,
\qquad
0u^k\leftrightarrow u^k0,
\qquad
00\leftrightarrow u^kd^k.
\end{equation}

 Beyond the height constraint, colorful Motzkin paths obey a strict last-in-first-out color-matching condition. Any downward step $d^k$ must exclusively close the most recent unmatched upward step $u^k$. A valid height profile is therefore a necessary but not sufficient condition for path legality. For instance, the state $\ket{u^1d^2}$ exhibits a valid height sequence $(1,0)$ yet corresponds to an illegal colorful path due to mismatched color ordering.

 To systematically track color matching information, we adopt a stack structure as the minimal causal memory unit and decompose the stack vector $q_i$ as 
\begin{equation}
  q_i=\bigoplus_{h=1} q_i^{h},\qquad q_i^{h}\in{0,\pm1,\ldots\pm s},
\end{equation}
where each component \(q_i^{(h)}\) encodes the color of the unmatched upward step residing at height \(h\). We further introduce \(\chi_i\) as a local color validity flag, such that \(\chi_i = 1\) if the current color configuration satisfies the last-in-first-out matching rule, and \(\chi_i = 0\) otherwise.

Let \(S_{i-1}\) denote the stack height prior to processing site \(i\).  The update rules  for the stack vector \(q_i\) and color flag \(\chi_i\) fall into three distinct cases:
\begin{description}
    \item[Push \((x_i=u^k)\)]  Set \(q_i^{S_{i-1}+1}=c_i\), inherit all remaining entries of \(q_{i-1}\) to form the full stack vector \(q_i\), and assign the local color flag \(\chi_i=1\): 

\item[Flat \((x_i=0)\)]  Preserve the stack state by setting \(q_i=q_{i-1}\) and \(\chi_i=1\):

\item[Pop \((x_i=d^k)\)]  If \(S_{i-1}=0\) or the color stored at the current stack top \(q_{i-1}^{S_{i-1}}\ne c_i\), leave the stack state unchanged and set \(\chi_i=0\). Otherwise, copy the full stack \(q_{i-1}\) to \(q_i\), erase the stack-top entry at height \(S_{i-1}\), and set \(\chi_i=1\):
\end{description}
Corresponding to these three cases, the stack vector $q_i$ is updated according to the equation
\begin{equation}
     q_i =
     \begin{cases}
         q_{i-1} \oplus c_i , & {\rm if} \, \, \Delta_i =1 , \\
       q_{i-1}/q^{S_{i-1}}_{i-1}  , & {\rm if} \, c_i = q_{i-1}^{S_{i-1}} , \, \Delta_i=-1 ,\\
         q_{i-1} , & {\rm otherwise} ,
     \end{cases}
     \label{Eq:Stack}
\end{equation}
where $q_{i-1}/q^{S_{i-1}}_{i-1}$ denotes the removal of the top element $q^{S_{i-1}}_{i-1}$ from the stack vector $q_{i-1}$. 
The color validity flag follows the update equation
\begin{equation}
    \chi_i = 
    \begin{cases}
    0 & {\rm if} \,\, \Delta_i=-1 , \,\, c_i \not= q_{i-1}^{S_{i-1}} ,,  \\
    1, & {\rm otherwise} .
    \end{cases}
    \label{Eq:ColorFlag}
\end{equation}

The ReLU color gate
\begin{equation}
v_i=\sigma(\chi_i)
\label{eq:stack_gate}
\end{equation}
is one iff the color operation at site \(i\) is stack-consistent.

\subsection{Stack-augmented RNN}
\label{subsec:RNN_colorful}

 Similar to the colorless case discussed in Sec.~\ref{sec:RNN_colorless}, RNN gives a direct method to construct a representation of a colorful Motzkin state. However, because of the extra color rule, the previous RNN structure can not satisfy our requirement. Therefore, we use a stack-augmented RNN structure, which contains a height RNN and a color RNN, to recurrently determine the height and color wavefunctions, respectively. 

 Analogous to the colorless formulation presented in Sec.~\ref{sec:RNN_colorless}, RNN provides a natural framework for constructing neural-network representations of colorful Motzkin states. However, the additional color-matching constraints invalidate the standard RNN architecture designed for colorless paths, which only enforces height consistency. To resolve this limitation, we adopt a stack-augmented dual-branch RNN structure, consisting of a height RNN and a color RNN. The two branches perform recurrent updates independently to characterize the height wavefunction and color configuration wavefunction, respectively.
 
 For the height branch, we inherit the variables $\Phi_i$ and $h_i$ defined in the colorless RNN framework (Sec.~\ref{sec:RNN_colorless}). Replacing the input $x_i$ in Eqs.~(\ref{eq:RNN_colorless_S}--\ref{eq:RNN_colorless_activation}) with the height increment $\Delta_i$ yields the feed-forward update pipeline for the height subnetwork
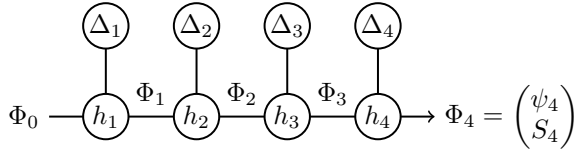
\[
\begin{tikzpicture}[every node/.style={scale=1},scale=0.6]
\draw (2,0) node {$\Delta_1$} circle (0.5)
      (4,0) node {$\Delta_2$} circle (0.5)
      (6,0) node {$\Delta_3$} circle (0.5)
      (8,0) node {$\Delta_4$} circle (0.5) ;

\draw (2,-2) node {$h_1$} circle (0.5)
      (4,-2) node {$h_2$} circle (0.5)
      (6,-2) node {$h_3$} circle (0.5)
      (8,-2) node {$h_4$} circle (0.5) ;

\draw (3,-1.5) node {$\Phi_1$}
      (5,-1.5) node {$\Phi_2$}
      (7,-1.5) node {$\Phi_3$};

\draw (2,-0.5) -- (2,-1.5);
\draw (4,-0.5) -- (4,-1.5);
\draw (6,-0.5) -- (6,-1.5);
\draw (8,-0.5) -- (8,-1.5);
\draw (0.75,-2) node [left] {$\Phi_0$} -- (1.5,-2);
\draw (2.5,-2) -- (3.5,-2);
\draw (4.5,-2) -- (5.5,-2);
\draw (6.5,-2) -- (7.5,-2);
\draw[->] (8.5,-2) -- (9.25,-2) [right] node  {$\Phi_4=\begin{pmatrix}
          \psi_4\\
          S_4
      \end{pmatrix}$};
\end{tikzpicture}
\]
The final output $\Phi_N=(\psi_N,S_N)^T$ (illustrated here for $N=4$) encodes the height validity indicator $\psi_N$, which quantifies whether the path satisfies the global height constraint.

Complementing the height branch, we design a structurally consistent RNN subnetwork to model color dynamics  
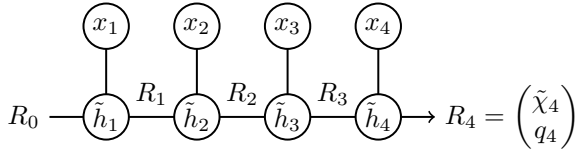
\[
\begin{tikzpicture}[every node/.style={scale=1},scale=0.6]
\draw (2,0) node {$x_1$} circle(0.5)
      (4,0) node {$x_2$} circle(0.5)
      (6,0) node {$x_3$} circle(0.5)
      (8,0) node {$x_4$} circle(0.5);

\draw (2,-2) node {$\tilde{h}_1$} circle(0.5)
      (4,-2) node {$\tilde{h}_2$} circle(0.5)
      (6,-2) node {$\tilde{h}_3$} circle(0.5)
      (8,-2) node {$\tilde{h}_4$} circle(0.5);

\draw (3,-1.5) node {$R_1$}
      (5,-1.5) node {$R_2$}
      (7,-1.5) node {$R_3$};

\draw (2,-0.5) -- (2,-1.5);
\draw (4,-0.5) -- (4,-1.5);
\draw (6,-0.5) -- (6,-1.5);
\draw (8,-0.5) -- (8,-1.5);
\draw (0.75,-2) node [left] {$R_0$} -- (1.5,-2);
\draw (2.5,-2) -- (3.5,-2);
\draw (4.5,-2) -- (5.5,-2);
\draw (6.5,-2) -- (7.5,-2);
\draw[->] (8.5,-2) -- (9.25,-2) [right] node {$R_4=\begin{pmatrix}
    \tilde{\chi}_4\\q_4
\end{pmatrix}$};
\end{tikzpicture}
\]
This color RNN evolves the composite color state vector
\begin{equation}
    R_i=\begin{pmatrix}
        \tilde{\chi}_i\\q_i
    \end{pmatrix},
\end{equation}
where $q_i$ is the color stack vector, $\chi_i$ denotes the local color validity flag and $\tilde{\chi}_i$ denotes the prefix color validity flag. The initial state is set to $R_0=(1,\emptyset)^T$. Each hidden neuron $\tilde{h}_i$ takes ($R_{i-1}, \Delta_i, c_i$) as input, it recursively updates the stack vector $q_i$ following Eq.~\eqref{Eq:Stack} with the local color validity flag $\chi_i$ following Eq.~\eqref{Eq:ColorFlag}, and global validity flag $\tilde{\chi}_i$ is generated by
\begin{equation}
    \tilde{\chi}_i = \tilde{\chi}_{i-1}\chi_i\,.
\end{equation}

 The update of $\tilde{\chi}_t$, which is equivalent to the product over all local color-validity flags, ensures that a color mismatch at any intermediate site permanently sets the wave-function amplitude to zero. Together with the height indicator $\psi_N$, the resulting wave function,
 \begin{equation}
  \Psi(X)
  =\psi_N\tilde{\chi}_N ,
\end{equation}
 satisfies $\Psi(X)=1$ iff $X$ is a valid colorful Motzkin path, and vanishes otherwise.

\subsection{Feedforward Neural Network}

 The last-in-first-out stack rule can be viewed as a causal pointer.  For a down step at site \(i\), its matching up step is the most recent preceding up step satisfying the height consistency condition \(S_{i-1}=S_i+1\), where the post-up height of the matched up step equals the pre-pop height of the target down step. Such paired up-down motion lies on the same horizontal level. This observation suggests that we can slightly modify the colorless FNN architecture to obtain an FNN representation for the colorful case.

 The proposed colorful FNN retains the hierarchical layer design of the original colorless framework. Specifically, one dedicated layer is employed to compute prefix sums identically to the colorless state, followed by \(N-1\) successive layers that encode the nearest preceding up-step information for each input site. A full schematic visualization of this network is presented below: 
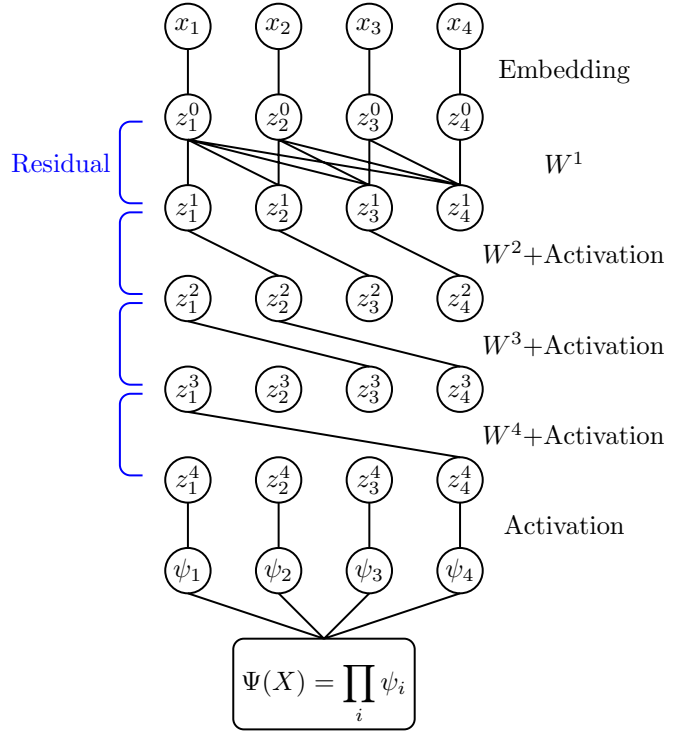
\[
\begin{tikzpicture}[every node/.style={scale=1},scale=0.6]
\draw (8.3,-1) node {Embedding};
\draw (8.3,-3) node {$W^1$};
\draw (8.5,-5) node {$W^2$+Activation};
\draw (8.5,-7) node {$W^3$+Activation};
\draw (8.5,-9) node {$W^4$+Activation};
\draw (8.3,-11) node {Activation};
\draw (0,0) node {$x_1$} circle (0.5)
      (2,0) node {$x_2$} circle (0.5)
      (4,0) node {$x_3$} circle (0.5)
      (6,0) node {$x_4$} circle (0.5) ;
\draw (0,-0.5)--(0,-1.5)
      (2,-0.5)--(2,-1.5)
      (4,-0.5)--(4,-1.5)
      (6,-0.5)--(6,-1.5);
\draw (0,-2) node {$z_1^0$} circle (0.5)
      (2,-2) node {$z_2^0$} circle (0.5)
      (4,-2) node {$z_3^0$} circle (0.5)
      (6,-2) node {$z_4^0$} circle (0.5) ;
\draw (0,-2.5) -- (0, -3.5) (0,-2.5) -- (2,-3.5) (0,-2.5) -- (4,-3.5) (0,-2.5) -- (6,-3.5)
      (2,-2.5) -- (2,-3.5) (2,-2.5) -- (4,-3.5) (2,-2.5) -- (6,-3.5)
       (4,-2.5) -- (4,-3.5) (4,-2.5) -- (6,-3.5)
       (6,-2.5) -- (6,-3.5) ;
\draw (0,-4) node {$z_1^1$} circle (0.5)
      (2,-4) node {$z_2^1$} circle (0.5)
      (4,-4) node {$z_3^1$} circle (0.5)
      (6,-4) node {$z_4^1$} circle (0.5) ;
\draw (0,-4.5)--(2,-5.5) (2,-4.5)--(4,-5.5) (4,-4.5)--(6,-5.5);
\draw (0,-6) node {$z_1^2$} circle (0.5)
      (2,-6) node {$z_2^2$} circle (0.5)
      (4,-6) node {$z_3^2$} circle (0.5)
      (6,-6) node {$z_4^2$} circle (0.5) ;
\draw (0,-6.5)--(4,-7.5) (2,-6.5)--(6,-7.5);
\draw (0,-8) node {$z_1^3$} circle (0.5)
      (2,-8) node {$z_2^3$} circle (0.5)
      (4,-8) node {$z_3^3$} circle (0.5)
      (6,-8) node {$z_4^3$} circle (0.5) ;
\draw (0,-8.5)--(6,-9.5);
\draw (0,-10) node {$z_1^4$} circle (0.5)
      (2,-10) node {$z_2^4$} circle (0.5)
      (4,-10) node {$z_3^4$} circle (0.5)
      (6,-10) node {$z_4^4$} circle (0.5) ;
\draw (2,-10.5)--(2,-11.5) (4,-10.5)--(4,-11.5) (0,-10.5)--(0,-11.5) (6,-10.5)--(6,-11.5);
\draw (0,-12) node {$\psi_1$} circle (0.5)
      (2,-12) node {$\psi_2$} circle (0.5)
      (4,-12) node {$\psi_3$} circle (0.5)
      (6,-12) node {$\psi_4$} circle (0.5) ;
\draw (2,-12.5)--(3,-13.5) (4,-12.5)--(3,-13.5) (0,-12.5)--(3,-13.5) (6,-12.5)--(3,-13.5);
\draw[rounded corners] (1,-15.5) rectangle (5,-13.5);
\draw (3,-14.7) node {$\Psi(X)= \displaystyle \prod_i \psi_i$};
\draw (-2.8,-3) node {\textcolor{blue}{Residual}};
\draw[draw=blue,rounded corners] (-1,-2.1)--(-1.5,-2.1)--(-1.5,-3.9)--(-1,-3.9);
\draw[draw=blue,rounded corners] (-1,-4.1)--(-1.5,-4.1)--(-1.5,-5.9)--(-1,-5.9);
\draw[draw=blue,rounded corners] (-1,-6.1)--(-1.5,-6.1)--(-1.5,-7.9)--(-1,-7.9);
\draw[draw=blue,rounded corners] (-1,-8.1)--(-1.5,-8.1)--(-1.5,-9.9)--(-1,-9.9);
\end{tikzpicture}
\]

We first embed each input configuration into a 4-dimensional feature vector
\begin{equation}
    z_i^k=(\Delta_i,c_i,S_i,n_i^k)^T,
\end{equation}
initialized as
\begin{equation}
 z_i^0=(\Delta_i,c_i,0,\sigma(x_i))^T.
\end{equation}
Here, \(n_i^k\) denotes the color attribute of the nearest preceding up step for the \(i\)-th site, captured up to the \(k\)-th layer, while the index \(k\) enumerates the hierarchical layers of the FNN.

As in the colorless FNN, the prefix sum of height increments is determined by the linear transformation 
\begin{equation}
    S_i=\sum_jW_{ij}^1\Delta_j, 
\end{equation}
where $W^1=W$ is the lower triangular matrix, defined in Eq. \eqref{eq:prefix_sum}. Together with the residual connection, this yields
have
\begin{equation}
  z_i^1=(\Delta_i,c_i,S_i,\sigma(x_i))^T.
\end{equation}

For the subsequent $2 \le k \le N$ layers (with $N=4$ here), we implement a layer-specific linear transformation:
\begin{equation}
    \left( p_i, d_i , b_i \right)  =  \sum_jW_{ij}^k \left(c_j , S_j, \Delta_j \right) , 
\end{equation}
where the auxiliary variables \(p_i\), \(d_i\), and \(b_i \) are intermediate features encoding color information, prefix sum values, and height increments, respectively. The layer-specific weight matrix obeys the binary rule
\begin{equation}
    W_{ij}^k = \delta_{j, i-k+1}
\end{equation}
This weight design enables the \(k\)th layer to extract \( (p_i, d_i, b_i) \) from the \(k\)th preceding site \((i-k+1)\). 

The layer-wise activation function is defined as
\begin{equation}
    dn_i^{k} = p_i\delta_{n_i^{k-1},0}\delta_{d_i,S_i+1}\delta_{b_i,1} .
\end{equation}
The update \(dn_i^k=p_i\) is activated iff the \((i+1-k)\)-th site is an up step horizontally aligned with the \(i\)th site, and no valid color is stored in \(n_i^{k-1}\) in prior layers. Under this condition, the updated feature \(n_i^{k-1}+dn_i^{k}\) stores the valid color information of the matched up step.

Note that the Kronecker delta functions in the activation can be fully implemented via rectified linear unit (ReLU) activations, through the identities
\begin{eqnarray}
    \delta_{x,y}&=&\sigma(1-|x-y|)\\
    |x|&=&\sigma(x)+\sigma(-x)
    \label{abs_delta_relu}
\end{eqnarray}

After applying residual connections, the updated feature vector reads
\begin{equation}
    z_i^{k}=(\Delta_i,c_i,S_i,n_i^{k-1}+dn_i^{k})^T .
\end{equation}
After propagating through all layers, the final output feature vector converges to
\begin{equation}
    z_i^N=(\Delta_i,c_i,S_i,n_i)^T,
\end{equation}
where \(n_i\) uniquely encodes the color information of the nearest valid preceding up step. The site-wise wavefunction \(\psi_i(X)\) is then computed via a composite activation function:
\begin{eqnarray}
    \psi_i(X)&=&y_i(z_i^N)\,v_i(z_i^N)\\
    y_i(z_i^N)&=&\sigma(1-\sigma(-S_i)-\sigma(i-N+1)\sigma(S_i))\\
    v_i(z_i^N)&=&(1-\delta_{\Delta_i,-1})+\delta_{n_i,c_i}\delta_{\Delta_i,-1}
\end{eqnarray}
Here, \(y_i\) enforces the height validity, while \(v_i\) imposes the color consistency constraint for each configuration.
Finally, we get the wavefunction $\Psi(X)=\prod_i \psi_i $ up to a normalization constant. 

One potential concern is that the color validity term \(v_i\) does not explicitly constrain the stack to be empty after \(N\) moves. This omission, however, does not affect the validity of the construction. A  non-empty stack will result in a non-zero final height, which is strictly penalized by the height validity product \(\prod_i y_i\). In such cases, the overall wavefunction amplitude vanishes.

 We note that an analogous modification applies to the colorful CNN construction, starting from its colorless counterpart. The first layer employs the same convolution kernel as in the colorless case, while for all subsequent layers \(2\le k\le N\), we employ one-hot convolution kernels defined by
\begin{equation}
    (\mathrm{Kernel}_k)_j=\delta_{j,N-k+1} , 
\end{equation}
 which reproduce the exact feature extraction operations of the corresponding FNN layers described above. Given the structural consistency and functional equivalence between the modified FNN and CNN for colorful cases, we do not present a separate section devoted to the CNN construction.

\subsection{Transformer}
\label{subsec:transformer_colorful}

 Beyond the ansatz built on FNN, CNN, and stack-augmented RNN, we propose an alternative construction leveraging causally masked self-attention. The design adopts a three-layer multi-head transformer equipped with hard softmax attention and pointwise multi-layer perception (MLP) modules:
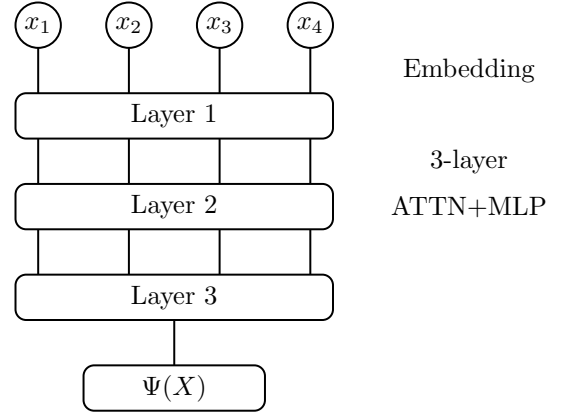
\[
\begin{tikzpicture}[every node/.style={scale=1},scale=0.6]
\draw (9.5,-1) node {Embedding};
\draw (9.5,-3) node {3-layer};
\draw (9.5,-4) node {ATTN+MLP};
\draw (0,0) node {$x_1$} circle (0.5)
      (2,0) node {$x_2$} circle (0.5)
      (4,0) node {$x_3$} circle (0.5)
      (6,0) node {$x_4$} circle (0.5) ;
\draw (0,-0.5)--(0,-1.5)
      (2,-0.5)--(2,-1.5)
      (4,-0.5)--(4,-1.5)
      (6,-0.5)--(6,-1.5);
\draw[rounded corners] (-0.5,-2.5) rectangle (6.5,-1.5);
\draw (3,-2) node{Layer 1};
\draw (0,-2.5)--(0,-3.5)
      (2,-2.5)--(2,-3.5)
      (4,-2.5)--(4,-3.5)
      (6,-2.5)--(6,-3.5);
\draw[rounded corners] (-0.5,-4.5) rectangle (6.5,-3.5);
\draw (3,-4) node{Layer 2};
\draw (0,-4.5)--(0,-5.5)
      (2,-4.5)--(2,-5.5)
      (4,-4.5)--(4,-5.5)
      (6,-4.5)--(6,-5.5);
\draw[rounded corners] (-0.5,-6.5) rectangle (6.5,-5.5);
\draw (3,-6) node{Layer 3};
\draw (3,-6.5)--(3,-7.5);
\draw[rounded corners] (1,-8.5) rectangle (5,-7.5);
\draw (3,-8) node{$\Psi(X)$};

\end{tikzpicture}
\]
The internal layer structure is illustrated below:
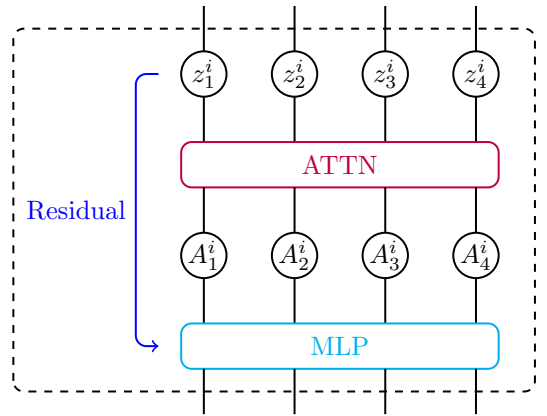
\[
\begin{tikzpicture}[every node/.style={scale=1},scale=0.6]
\draw (0,-0.5)--(0,-1.5)
      (2,-0.5)--(2,-1.5)
      (4,-0.5)--(4,-1.5)
      (6,-0.5)--(6,-1.5);
\draw (0,-2) node {$z_1^{i}$} circle (0.5)
      (2,-2) node {$z_2^{i}$} circle (0.5)
      (4,-2) node {$z_3^{i}$} circle (0.5)
      (6,-2) node {$z_4^{i}$} circle (0.5);
\draw (0,-2.5)--(0,-3.5)
      (2,-2.5)--(2,-3.5)
      (4,-2.5)--(4,-3.5)
      (6,-2.5)--(6,-3.5);
\draw [rounded corners,draw=purple] (-0.5,-4.5)  rectangle (6.5,-3.5);
\draw (3,-4) node {\textcolor{purple}{ATTN}};
\draw (0,-4.5)--(0,-5.5)
      (2,-4.5)--(2,-5.5)
      (4,-4.5)--(4,-5.5)
      (6,-4.5)--(6,-5.5);
\draw (0,-6) node {$A_1^{i}$} circle (0.5)
      (2,-6) node {$A_2^{i}$} circle (0.5)
      (4,-6) node {$A_3^{i}$} circle (0.5)
      (6,-6) node {$A_4^{i}$} circle (0.5);
\draw (0,-6.5)--(0,-7.5)
      (2,-6.5)--(2,-7.5)
      (4,-6.5)--(4,-7.5)
      (6,-6.5)--(6,-7.5);
\draw [rounded corners,draw=cyan] (-0.5,-8.5)  rectangle (6.5,-7.5);
\draw (3,-8) node {\textcolor{cyan}{MLP}};
\draw (0,-8.5)--(0,-9.5)
      (2,-8.5)--(2,-9.5)
      (4,-8.5)--(4,-9.5)
      (6,-8.5)--(6,-9.5);
\draw[->,draw=blue,rounded corners] (-1,-2)--(-1.5,-2)--(-1.5,-8)--(-1,-8);
\draw(-1.5,-5) node[left] {\textcolor{blue}{Residual}};
\draw[dashed, rounded corners] (-4.2,-9) rectangle(7.3,-1);
\end{tikzpicture}
\]
 At initialization, we embed each input configuration $x_t$ into a 9-dimensional feature vector 
\begin{equation}
    z_t^0=(x_t,t,\Delta_t,0,0,0,0,0,1)^T .
\end{equation}

We now detail the function of each layer in sequence.

{\bf Layer 1: prefix sums and height constraints.} We set the projection weights
\begin{equation}
    W_Q^1=W_K^1=0, \quad W_V^1=e_1e_3^T, 
\end{equation}
where $e_1$ and $e_3$ are the first and third one-hot basis in the embedded vector space, respectively. This yields,   
\begin{eqnarray}
    W_V^1z_t^0=(\Delta_t,0,0,0,0,0,0,0,0)^T
\end{eqnarray}
    
and a lower triangular causal mask matrix
\begin{eqnarray}
    M_{ti}&=&
    \begin{cases}
    0 & t\ge i\\
    -\infty& t<i
    \end{cases}.
\end{eqnarray}
The resulting attention logit reads
\begin{eqnarray}
    \mathrm{Logit}^1(t,i)&=&M_{ti}+(W_Q^1 z_t^0)^T(W_K^1z_i^0)=M_{ti}\,.
\end{eqnarray}
Following the weighting construction introduced in Eq.~\eqref{eq:attn_alpha}, we obtain the attention weights
\begin{eqnarray}
    \alpha(t,i)&=&
    \begin{cases}
       \displaystyle 1/t , & t\ge i ,\\
        0 , & t<i .
    \end{cases}
\end{eqnarray}

Applying the averaging procedure of Eq.~\eqref{eq:prefix_avg}, we obtain $A_t^1 = S_t/t$. The subsequent MLP computes
\begin{eqnarray}
    S_t&=&t\cdot A_t^1 =S_t ,\\
    S_{t-1}&=&S_t-\Delta_t ,\\
    B_t&=&\sigma(-S_t) .
\end{eqnarray}
Incorporating the residual connection yields 
\begin{equation}
    z_t^1=\big(x_t,t,\Delta_t,S_t,S_{t-1},(S_{t-1})^2,B_t,0,1\big)^T .
\end{equation}

{\bf Layer 2: color matching.} The projection operators act as  
\begin{eqnarray}
W_Q^2z_t^1&=&
\begin{pmatrix}
S_t\\
1
\end{pmatrix}\\
W_K^2z_t^1&=& \begin{pmatrix}
2N\omega S_{t-1}\\
-N\omega (S_{t-1})^2 +\omega\cdot t
\end{pmatrix}\\
W_V^2z_t^1&=& x_t\,.
\end{eqnarray}
 Here we introduce a parameter $\omega$ satisfying  $ \omega \gg 1$, where $N\omega$ and $\omega$ act as filters for height and horizontal distance, respectively. This hierarchy of scales is selected to ensure the distance filter does not interfere with the height filter. Both filters decay rapidly as a function of the site separation. The corresponding attention logit becomes 
\begin{eqnarray}
    &&\mathrm{Logit}^2(t,i) \nonumber \\
    &=&M_{t,i}+(W_Q^2z_t^1)^T (W_K^2z_i^1)\notag\\
    &=&M_{t,i}-\beta (S_t -S_{i-1})^2 +\beta (S_t)^2+\omega \cdot i\,.
\end{eqnarray}
For fixed $t$, the term $\beta (S_t)^2$ shifts the logits uniformly across all $i$, and therefore cancels out upon softmax normalization.

The output of this attention block satisfies
\begin{eqnarray}
    A_t^2 =x_{j}
\end{eqnarray}
for every down step at $t$, where $j$ is the most recent preceding up step sharing the same height as site $t$. We note that when no such prior up step exists, attention weights become nearly uniform and $A_t^2$ may return an arbitrary color.  This, however, does not affect the wavefunction since such configurations necessarily violate the height constraint and correspond to invalid paths. The MLP then evaluates
\begin{equation}
    \Gamma_t =\delta(x_t<0)\cdot \delta(x_t+A_t^2\ne 0) ,
\end{equation} 
which flags a violation of the color-matching rule. With the residual connection, we obtain 
\begin{equation}
    z_t^2=\big(x_t,t,\Delta_t,S_t,S_{t-1},(S_{t-1})^2,B_t,\Gamma_t,1\big)^T .
\end{equation}

{\bf Layer 3: global aggregation and readout.} We set 
\begin{equation}
    W_Q^3=W_K^3=0, \quad  W_V^3 z_t^2=(B_t,\Gamma_t)^T. 
\end{equation}
The attention output takes the form
\begin{eqnarray}
    A_t^3=\frac{1}{t}\sum_{i=1}^t\begin{pmatrix}
        {B_i}\\
        {\Gamma_i}
    \end{pmatrix}=\begin{pmatrix}
        A_t^3[1]\\
        A_t^3[2]
    \end{pmatrix}\,.
\end{eqnarray}
Consequently, $A_N^3$ aggregates all the constraint violation statistics along the full path. The final MLP relies solely on $A_N^3$ to produce the wavefunction 
\begin{eqnarray}
    y_N&=&\sigma(1-|S_N|-N\cdot A_N^3[1])\\
    v_N&=&\sigma(1-N\cdot A_N^3[2])\\
    \Psi(X)&=&\sigma(y_N+v_N-1)=y_Nv_N\,.
\end{eqnarray}
Here $y_N$ and $v_N$ encode the height and color criteria, respectively. The resulting wavefunction $\Psi(X)=1$ iff $y_N=v_N=1$ and 0 otherwise. 

The transformer construction has a more complex architecture than the other approaches, which may seem excessive for representing Motzkin states alone. However, it offers the potential to be extended into a variational NQS for studying quantum systems related to the Motzkin model that are not exactly soluble.

\section{Discussion}
\label{sec:discussion}

\subsection{Parameter complexity of neural-network representations}

 We have presented exact neural-network representations for both critical and supercritical Motzkin states. The colorless formulation confines the wavefunction support to prefix-sum constraints, supplemented by nonnegativity and endpoint conditions. The colorful construction further imposes a last-in-first-out color-matching rule. In all cases, the network weights are fixed analytically by the Motzkin legality rules. The different representations, however, involve distinct parameter counts, whose scaling with the system size $N$ is summarized and compared in Table~\ref{tab:representation_comparison}, along with their tensor-network counterparts.

 The RNN construction uses few parameters, of order \(\mathcal{O}(1)\) for the colorless height rule and \(\mathcal{O}(N)\) for an explicit color stack. Its sequential update, however, is not naturally parallel, and the colorful stack operations are not differentiable. The RNN is therefore best viewed as an algorithmic template rather than a practical training ansatz.

  The FNN construction encodes prefix sums through dense triangular maps. It requires an \(N\times N\) weight matrix for the colorless state and \(N\) such matrices for the colorful pointer construction, giving \(\mathcal{O}(N^2)\) and \(\mathcal{O}(N^3)\) scaling, respectively. Because the colorful pointer is equivalent to the stack rule but uses only differentiable operations, the FNN and the subsequent architectures are computationally trainable. 

  The CNN construction replaces the dense \(N\times N\) weight matrices by convolution kernels of length \(N\). Prefix sums are then generated by convolution rather than by a full linear transform, which reduces the parameter count to \(\mathcal{O}(N)\) for the colorless state and \(\mathcal{O}(N^2)\) for the colorful one.

The colorful transformer employs three \(9\times 9\) projection matrices, \(W_Q\), \(W_K\), and \(W_V\), which act on the embedded vectors in each layer. Since the mask is fixed rather than variational in the general transformer construction, the total number of variational parameters scales as \(\mathcal{O}(1)\).

The colorless transformer construction presented in Sec.~\ref{sec:attention} has a simple and intuitive structure, but is not optimal in terms of parameter scaling. Because the embedded vectors have dimension \(N+2\), each of the three projection matrices \(W_Q\), \(W_K\), and \(W_V\) has dimensions \((N+2)\times(N+2)\). Consequently, the number of parameters scales as \(\mathcal{O}(N^2)\). This scaling can nevertheless be reduced to \(\mathcal{O}(1)\) by implementing the colorless transformer within the colorful architecture presented in Sec.~\ref{subsec:transformer_colorful} and setting \(s=1\).

\begin{table}[!t]
\caption{Parameter scaling versus chain length $N$ for various representations of colorless and colorful Motzkin states. Results for the tensor-network-state (TNS) representation are shown for comparison.}
\label{tab:representation_comparison}
\renewcommand{\arraystretch}{1.18}
\begin{ruledtabular}
\begin{tabular}{ccc}
Representation & Colorless & Colorful \\
\hline
RNN & \(\mathcal{O}(1)\) & \(\mathcal{O}(N)\) \\
FNN & \(\mathcal{O}(N^2)\) & \(\mathcal{O}(N^3)\) \\
CNN & \(\mathcal{O}(N)\) & \(\mathcal{O}(N^2)\) \\
Transformer & \(\mathcal{O}(1)\) & \(\mathcal{O}(1)\) \\
TNS & \(\mathcal{O}(N\log_2 N)\)~\cite{alexander2021exact}
& \(\mathcal{O}((2s+2)^4N^2)\)~\cite{alexander2019exact} \\
\end{tabular}
\end{ruledtabular}
\end{table}

  All colorful constructions can represent the colorless state without structural modification. The colorless Motzkin state is the \(s=1\) specialization of the colorful problem. Once the height rule is enforced, the color-matching constraint becomes vacuous. Consequently, the colorful architectures remain exact for \(s=1\) if the color branch is retained but never activated, or equivalently if the color labels are fixed to a single value. The price is that one carries the full colorful parameter budget even though a dedicated colorless network would suffice. The scalings in the colorful column of Table~\ref{tab:representation_comparison} therefore also upper-bound the cost of representing the colorless state.

    Across these architectures, the parameter counts above refer to fixed analytic weights rather than to a trained variational ansatz. All of these weights are determined by \((N,s)\) and by the exact masks that encode Motzkin legality. They are not variationally optimized.

\subsection{Comparison with tensor-network representation}

 Exact tensor-network representations of Motzkin states have been constructed from elementary building-block tensors~\cite{alexander2019exact,alexander2021exact}. For the colorless Motzkin state, an exact hierarchical construction~\cite{alexander2021exact} uses \(\mathcal{O}(N\log_2 N)\) blocks of shape \(3\times 3\times 3\times 3\), so the native tensor-network parameter count scales as \(\mathcal{O}(N\log_2 N)\). For the colorful model~\cite{alexander2019exact}, one needs \(\mathcal{O}(N^2)\) building blocks of shape \((2s+2)\times(2s+2)\times(2s+2)\times(2s+2)\), giving a parameter count \(\mathcal{O}(s^4N^2)\). Relative to these native TNS costs, the comparison is architecture dependent. For the colorless state, the TNS scaling \(\mathcal{O}(N\log_2 N)\) is higher than RNN and CNN, but lower than the FNN and transformer budgets. For the colorful state, several neural-network constructions remain \(\mathcal{O}(N)\) or \(\mathcal{O}(N^2)\), matching or improving upon the TNS scaling. Beyond raw parameter counting, the neural constructions offer a more direct route to generalizations such as perturbed Motzkin models.

  The two frameworks are complementary rather than competing. Tensor networks encode Motzkin states through virtual spaces associated with height and color sectors, making the Schmidt structure across a cut explicit. This viewpoint underlies exact rainbow and holographic constructions~\cite{alexander2019exact,alexander2021exact}. The neural constructions instead treat the same states as recognition problems. A causal network reads the spin configuration, updates a small set of program variables, and returns a legality indicator. Tensor networks emphasize on the entanglement structure, while neural networks expose the computational structure of the support.

  This distinction matters for variational many-body physics. Neural-network ans\"atze are often optimized as black-box function approximators, whereas exact constructions identify which architectural motifs encode which physical constraints. For Motzkin states, prefix aggregation is the natural primitive for the height rule, and a stack or attention pointer is the natural primitive for nested colors. Analogous motifs appear in other constrained states, for example, Dyck and Fredkin paths require parenthesis matching, gauge-theory wave functions require local Gauss-law constraints, stabilizer states require parity checks, and string-net or loop-gas states require global connectivity information. Therefore, the present work suggests a general route to exact or near-exact neural-network states for a wider class of quantum states. 
  
  The same perspective clarifies why anomalous entanglement need not preclude a compact neural representation. The colorful Motzkin state has \(\Theta(\sqrt{N})\) half-chain entanglement entropy because many color strings can cross a cut, yet the legality of a complete configuration is still decided by a polynomial-size causal computation. In this sense, neural- and tensor-networks diagnose different kinds of complexity. Tensor networks are sensitive to bipartite entanglement, whereas causal neural networks can exploit algorithmic structure in the computational basis. This complementarity is potentially useful for area-deformed Motzkin models~\cite{zhang2017novel,levine2017gap}, Fredkin chains~\cite{salberger2017fredkin}, and Rydberg implementations of constrained spin-1 dynamics~\cite{mukherjee2026rydberg}, where MPS bond dimensions can become large while the physically allowed configurations remain highly structured.

  More broadly, Motzkin states show that exact neural-network representations can be obtained by design rather than by training. The fixed weights derived here provide benchmarks for studying optimization, sampling, and generalization in neural quantum states. This offers templates for exact neural-network representations of other quantum states whose amplitudes are governed by interpretable combinatorial rules.

\section{Summary}
\label{sec:summary}

 We have constructed exact neural-network representations of both colorless and colorful Motzkin states. The constructions are fully analytic, and all weights are fixed by the system size and the Motzkin legality rules. For the colorless state, a causal prefix-sum block, realized as a recurrent update, a dense triangular map, a one-sided convolution, or a masked attention layer, computes the path heights, while position-selective ReLU gates enforce nonnegativity at every prefix and return to zero at the endpoint. For the colorful state, this height block is augmented by a last-in-first-out color block, implemented either as an explicit stack or as an attention pointer to the matching up step, which enforces the nested color-matching rule.

 Despite the anomalous entanglement of the target states, logarithmic for the colorless state and \(O(\sqrt{N})\) for the colorful state, the resulting parameter counts remain polynomial. Depending on the architecture, they scale from \(\mathcal{O}(1)\) to \(\mathcal{O}(N^2)\) in the colorless case and from \(\mathcal{O}(N)\) to \(\mathcal{O}(N^3)\) in the colorful case, and are competitive with the exact tensor-network representations. 

 These results show that NQS can exactly represent combinatorially constrained wavefunctions beyond the entanglement area-law. More importantly, these constructions reveal a general design principle: when the support and phase structure of a many-body state can be computed by a finite causal algorithm, the algorithm may, under suitable architectural assumptions, be compiled into a faithful neural-network representation of the state. The fixed-weight architectures constructed herein provide analytically controlled benchmarks for optimization, sampling, and generalization. Furthermore, they serve as prototypes for investigating related constrained systems, including area-deformed Motzkin models~\cite{zhang2017novel,levine2017gap}, Fredkin spin chains~\cite{salberger2017fredkin}, and Rydberg simulator realizations of Motzkin-type dynamics~\cite{mukherjee2026rydberg}.

\vspace{1cm}

\begin{acknowledgments}
 This work was supported by the National Key Research and Development Project of China (Grants No.~2024YFA1408604, No.~2021ZD0301800, and No.~2022YFA1403900), the National Natural Science Foundation of China (Grants No.~12488201, No.~12322403, and No.~12347107), and the Strategic Priority Research Program of Chinese Academy of Sciences (Grant No.~XDB0500202).
\end{acknowledgments}

\section*{Code Availability}
The source code used to verify the neural-network constructions at small system sizes is publicly available~\cite{motzkin_nqs_code}.

\bibliography{Motzkin_NQS}

\end{document}

%% file: tikz_figs/motzkin_three_paths.tex
\begin{tikzpicture}[every node/.style={scale=1},scale=0.6,line cap=round,line join=round]
\tikzset{
  gridline/.style={draw=gray},
  pathline/.style={line width=1.5pt},
  redstep/.style={pathline,color={rgb,255:red,208;green,2;blue,27}},
  bluestep/.style={pathline,color={rgb,255:red,74;green,144;blue,226}},
  flatstep/.style={pathline,black}
}

\begin{scope}[shift={(0,14)}]
\node[anchor=west] at (-1.5,4.8) {(a)};
\node at (8.5,4.5) {crossing \(x\) axis};
\draw[draw=gray,fill=red!35] (7,0) rectangle (10,4);
\foreach \y in {0,...,4} {\draw[gridline] (0,\y)--(12,\y);}
\foreach \x in {0,...,12} {\draw[gridline] (\x,0)--(\x,4);}
\draw
  (-0.5,0) node {\(-1\)}
  (-0.5,1) node {\(0\)}
  (-0.5,2) node {\(1\)}
  (-0.5,3) node {\(2\)}
  (-0.5,4) node {\(3\)}
  (-1,2) node {\(y\)};
\foreach \x in {0,...,12} {\node at (\x,-0.5) {\(\x\)};}
\node at (6,-1) {\(x\)};
\draw[flatstep]
  (0,1)--(1,2)--(2,3)--(3,3)--(4,2)--(5,2)--(6,1)--(7,1)--(8,0)--(9,0)
  --(10,1)--(11,2)--(12,1);
\fill (0,1) circle (0.1) (12,1) circle (0.1);
\end{scope}

\begin{scope}[shift={(0,7)}]
\node[anchor=west] at (-1.5,4.8) {(b)};
\foreach \y in {0,...,4} {\draw[gridline] (0,\y)--(12,\y);}
\foreach \x in {0,...,12} {\draw[gridline] (\x,0)--(\x,4);}
\draw
  (-0.5,0) node {\(0\)}
  (-0.5,1) node {\(1\)}
  (-0.5,2) node {\(2\)}
  (-0.5,3) node {\(3\)}
  (-0.5,4) node {\(4\)}
  (-1,2) node {\(y\)};
\foreach \x in {0,...,12} {\node at (\x,-0.5) {\(\x\)};}
\node at (6,-1) {\(x\)};
\draw[flatstep] (0,0)--(1,0) (2,1)--(3,1) (7,3)--(8,3) (9,2)--(10,2);
\draw[redstep] (1,0)--(2,1) (5,1)--(6,2) (10,2)--(11,1)--(12,0);
\draw[bluestep] (3,1)--(4,2)--(5,1) (6,2)--(7,3) (8,3)--(9,2);
\fill (0,0) circle (0.1) (12,0) circle (0.1);
\end{scope}

\begin{scope}
\node[anchor=west] at (-1.5,4.8) {(c)};
\node at (7.5,5) {color mismatch};
\foreach \y in {0,...,4} {\draw[gridline] (0,\y)--(12,\y);}
\foreach \x in {0,...,12} {\draw[gridline] (\x,0)--(\x,4);}
\draw
  (-0.5,0) node {\(0\)}
  (-0.5,1) node {\(1\)}
  (-0.5,2) node {\(2\)}
  (-0.5,3) node {\(3\)}
  (-0.5,4) node {\(4\)}
  (-1,2) node {\(y\)};
\foreach \x in {0,...,12} {\node at (\x,-0.5) {\(\x\)};}
\node at (6,-1) {\(x\)};
\draw[flatstep] (0,0)--(1,0) (2,1)--(3,1) (7,3)--(8,3) (9,2)--(10,2);
\draw[redstep] (1,0)--(2,1) (5,1)--(6,2) (10,2)--(11,1)--(12,0) (8,3)--(9,2);
\draw[bluestep] (3,1)--(4,2)--(5,1) (6,2)--(7,3);
\fill (0,0) circle (0.1) (12,0) circle (0.1);
\draw[->] (6.5,4.5)--(6.5,2.6);
\draw[->] (8.5,4.5)--(8.5,2.6);
\end{scope}
\end{tikzpicture}